\def\year{2019}\relax
\documentclass[sigconf]{aamas} 
\usepackage{times}  
\usepackage{helvet}  
\usepackage{courier}  
\usepackage{url}  
\usepackage{graphicx}  
\usepackage{algorithm}
\usepackage{algpseudocode}
\usepackage{subcaption}
\usepackage{color}
\usepackage{tabu}
\usepackage{amsmath} 
\usepackage{amssymb}  
\usepackage{mathtools}
\usepackage{cleveref}
\usepackage{xcolor}
\usepackage{enumitem}

\setcopyright{ifaamas}  
\acmDOI{doi}  
\acmISBN{}  
\acmConference[AAMAS'19]{Proc.\@ of the 18th International Conference on Autonomous Agents and Multiagent Systems (AAMAS 2019), N.~Agmon, M.~E.~Taylor, E.~Elkind, M.~Veloso (eds.)}{May 2019}{Montreal, Canada}  
\acmYear{2019}  
\copyrightyear{2019}  
\acmPrice{}  

\begin{document}
\title[Collaborative Behaviors for Teams of Robotic Bodyguards]{Emergence of Scenario-Appropriate Collaborative Behaviors for Teams of Robotic Bodyguards}

\author{Hassam Ullah Sheikh}
\affiliation{%
 \institution{University of Central Florida}
 \city{Orlando}
 \state{Florida}
}
\email{hassam.sheikh@knights.ucf.edu}

\author{Ladislau B{\"o}l{\"o}ni}
\affiliation{%
 \institution{University of Central Florida}
 \city{Orlando}
 \state{Florida}
}
\email{lboloni@cs.ucf.edu}

\begin{abstract}

We are considering the problem of controlling a team of robotic bodyguards protecting a VIP from physical assault in the presence of neutral and/or adversarial bystanders in variety of scenarios. This problem is challenging due to the large number of active entities with different agendas and dynamic movement patterns, the need of cooperation between the robots as well as the requirement to take into consideration criteria such as social norms  in addition to the main goal of VIP safety.

In this paper we show how a multi-agent reinforcement learning approach can evolve behavior policies for teams of robotic bodyguards that outperform hand-engineered approaches. Furthermore, we propose a novel multi-agent reinforcement learning algorithm inspired by universal value function approximators that can learn policies which exhibit appropriate, distinct behavior in environments with different requirements.

\end{abstract}


\maketitle
\section{Introduction}
\label{sec:Introduction}


Recent progress in the field of autonomous robotics makes it feasible for robots to interact with multiple humans in public spaces. In this paper, we are considering a practical problem where a human VIP moving in various crowded scenarios is protected from physical assault by a team of robotics bodyguards. This problem has been previously explored in~\cite{Bhatia-2016-FLAIRS} where explicitly programmed behaviors of robots were used to carry out the task.



With the recent advancements in the single agent Deep RL~\cite{Mnih-2015-Nature, Silver-2016-Nature}, there has been a renewed interest in multi-agent reinforcement learning (MARL)~\cite{Shoham-2007-AI,Busoniu-2008-CSM}. Despite having outstanding performance in multiplayer games like Dota 2~\cite{OpenAI-2018} and Quake III Capture-the-Flag~\cite{Max-2018-ARXIV}, MARL algorithms have failed to learn policies that can work in different scenarios~\cite{Cruz-2014-SMC}. 



Providing physical protection to a VIP through robot bodyguards is a complex task where the robots must take into account the position and movement of the VIP, the bystanders and other robots. The variety of environments and scenarios in which the bodyguards need to act presents another challenge. We aim to solve the VIP protection problem through multi-agent deep reinforcement learning while simultaneously learning to communicate and coordinate between the robots. We propose a novel general purpose technique that allows multi-agent learners to learn distributed policies not only over the state space but also over a set of scenarios. We show that our solution outperforms a custom designed behavior, the quadrant load balancing method~\cite{Bhatia-2016-FLAIRS}. 



\section{The VIP Protection Problem}
\label{sec:ModelingBodyguard}

We are considering a VIP moving in a crowd of bystanders $\mathcal{B}=\left\{ b_{1},b_{2},\ldots,b_{m}\right\}$ protected from assault by a team of robot bodyguards $R=\left\{ r_{1},r_{2},\ldots,r_{n}\right\}$. To be able to reason about this problem, we need to quantify the {\em threat} to the VIP at a given moment - the aim of the bodyguards is to reduce this value.



Using the threat model defined in~\cite{Bhatia-2016-FLAIRS}, the residual threat $RT$ is defined as the threat to the VIP at time $t$ from bystanders $\mathcal{B}$. The {\em cumulative residual threat} to the VIP from bystanders $\mathcal{B}$ in the presence of bodyguards $R$ over the time period $[0,T]$ is defined as:
\begin{equation}
\label{eq:crt}
\mathrm{CRT}=\intop_{0}^{T}1-\prod_{i=1}^{k}\left(1-RT\left(VIP,b_{i}, R\right)\right)dt
\end{equation}
Our end goal is to minimize $\mathrm{CRT}$ through multi-agent reinforcement learning. Moreover~\cref{eq:crt} also forms the basis of our reward function for the VIP protection problem.

\section{Multi-Agent Universal Policy Gradient}


To solve the VIP protection problem under various scenarios, we propose {\em multi-agent universal policy gradient}: a multi-agent deep reinforcement learning algorithm that learns distributed policies not only over state space but also over a set of scenarios.  


Our approach uses Universal Value Function Approximators~\cite{Schaul-2015-ICML} to train policies and value functions that take a state-scenario pair as input. The outcome are a universal multi-agent policies that are able to perform on multiple scenarios as compared to policies that are trained and tested separately.

The main idea is to represent the different value function approximators for each agent $i$ by a single unified value function approximator that generalizes of over both state space and the scenarios. For agent $i$ we consider $V_i\left(s, g; \phi\right)\approx V_{i_g}^*\left(s\right)$ or
$Q_i\left(s, a, g; \phi\right)\approx Q_{i_g}^*\left(s, a\right)$ that approximate the optimal unified value functions over multiple scenarios and a large state space. These value functions can be used to extract policies implicitly or as critics for policy gradient methods.  We extend the idea of MADDPG~\cite{Lowe-2017-NIPS} with universal functional approximator, specifically we augment the centralized critic with scenario. 
Concretely, consider $N$ agents with policies $\pmb{\pi}=\{\pmb{\pi_1}, \ldots, \pmb{\pi_N}\}$ parameterized by $\pmb{\theta}=\{\pmb{\theta_1}, \ldots, \pmb{\theta_N}\}$ learning polices over $\pmb{G}$ scenarios the \textit{multi-agent universal policy gradient} for agent $i$ can written as



\begin{equation}
\label{eq:maupg}
 \nabla J_{\theta_{i}} = \mathbb{E}_{s,a, g\thicksim \mathcal{D}}\Bigg[\nabla_{\theta_{i}}\pi_{i}\left(a_i|o_{i}, g \right) \nabla_{a_i}Q_{i}^{\pi}\left(s,a_{1},\ldots,a_{N}, g\right)  \Bigg]
\end{equation}

\noindent where $s=\left(o_1, \ldots, o_N\right)$, $Q_{i}^{\pi}\left(s,a_{1},\ldots,a_{N}, g\right)$ is a centralized action-value function that takes the actions of all the agents, the state of the environment and the scenario to estimate the Q-value for agent $i$, ${a_i=\pi_i\left(o_i, g\right)}$ is action from agent $i$ following policy $\pi_i$ in scenario $g$ and $\mathcal{D}$ is the experience replay buffer.
\section{Experiments}
To investigate the effective of our proposed solution, we designed four scenarios inspired from possible real world situations of VIP protection and implemented them as behaviors in the Multi-Agent Particle Environment~\cite{Mordatch-2017-ARXIV}.
In each scenario, the participants are the VIP, four robot bodyguards and one or more classes of bystanders. The scenario description contains a number of {\em landmarks}, points on a 2D space that serve as starting point and destinations for the goal-directed movement by the agents. For each scenario, the VIP (brown disk) starts from the starting point and moves towards the destination landmark (green disk). The VIP exhibits a simple path following behavior, augmented with a simple social skill metric: it is about to enter the personal space of a bystander, it will slow down or come to a halt.



\begin{enumerate}[label=(\Alph*)]
\item \textbf{Random Landmark:} In this scenario, landmarks are placed randomly in the area. The bystanders are performing random waypoint navigation: they pick a random landmark, move towards it, and when they reached it, they choose a new destination.


\item \textbf{Shopping Mall:} In this scenario, landmarks are placed in fixed position on the periphery of the area, representing shops in a market. The bystanders visit randomly selected shops.

\item \textbf{Street:} The bystanders are moving towards waypoints that are outside the current area. However, due to their proximity to each other, the position of the other bystanders influence their movement described by laws of particles motion~\cite{Vicsek-1995-PRL}.


\item \textbf{Pie-in-the-Face}: In this ``red carpet'' scenario the one bystander take an active interest in the VIP. The {\em Unruly} bystander break the limit imposed by the line and try to approach the VIP (presumably, to throw a pie in his/her face).


\end{enumerate}


The observation of each agent is the physical state of the closest five bystanders in the environment and verbal utterances of all the agents $o_{i}=\left[x_{j,\ldots 5}, c_{k,\ldots N}\right]\in \mathcal{O}_i$ where $x_{j}$ is the observation of the entity $j$ from the perspective of agent $i$ and $c_{k}$ is the verbal utterance of the agent $k$. The scenario $g$ is represented as a one hot vector.

\section{Results}

\begin{figure}[H]
    \centering
    \includegraphics[height=1.9in, width=3.4in]{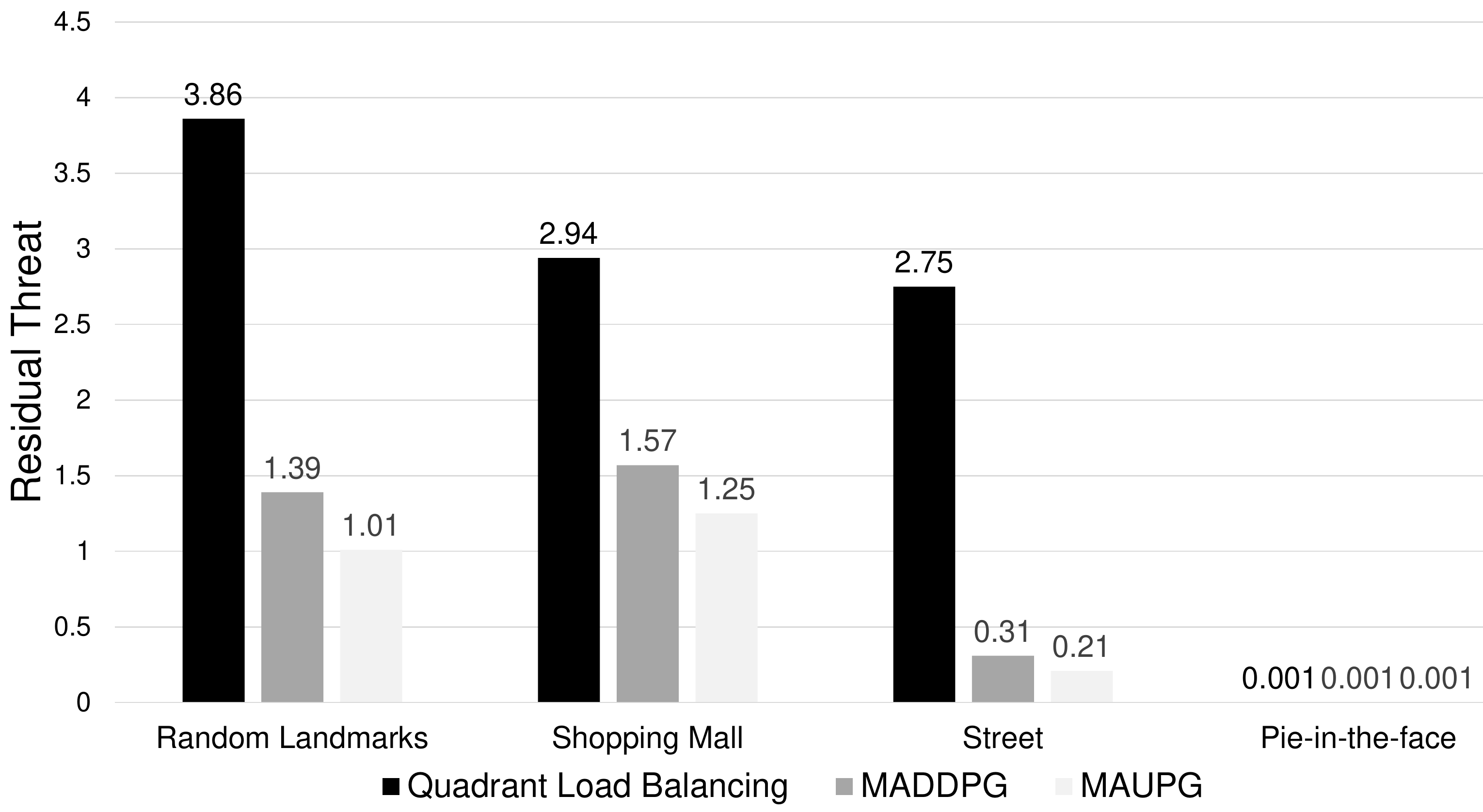}
    \caption{Comparing the average residual threat values for universal policy agents with MADDPG and QLB agents}
    \label{fig:combined}   
\end{figure}

In order to verify the claim that MARL algorithms trained on specific scenario fail to generalize over different scenarios, we evaluate policies trained via MADDPG on specific scenario and tested them on different scenarios. 
\begin{figure}[H]
        \centering
        \includegraphics[height=1.75in, width=3.40in]{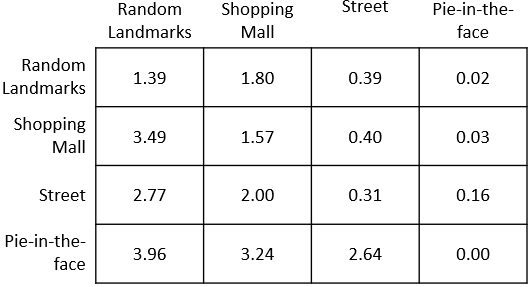}
\caption{A confusion matrix representing the average residual threat values of MADDPG policies trained on specific scenario when tested on different scenarios over 100 episodes.}
\label{tbl:cm}
\end{figure}
\vspace{-6pt}
\noindent From the results shown in~\Cref{tbl:cm} we can see that MADDPG policies trained on specific scenarios performed poorly when tested on different scenarios as compared to when tested on same scenario with different seeds.
In order to tackle the generalization problem, we train the agents using multi-agent universal policy gradient and compare with the results of scenario-dependant MADDPG policies and quadrant load balancing(QLB): a hand engineered technique to solve the VIP protection problem.The results can be seen in ~\Cref{fig:combined}. 
\section{Conclusions}
\label{sec:Conclusions}
In this paper, we highlighted the generalization problem faced by multi-agent reinforcement learning across different scenarios. To solve that problem we presented a novel algorithm that not only generalizes over state space but also over different scenarios. Using our solution, we solved the problem of providing physical protection to a VIP moving in a crowded space that outperforms another state-of-the-art multi-agent reinforcement learning algorithm as well as quadrant load-balancing: a hand engineered technique to solve the VIP protection problem. 

{\bf Acknowledgement:}  This research was sponsored by the Army
Research Laboratory and was accomplished under Cooperative Agreement Number
W911NF-10-2-0016. The views and conclusions contained in this document are
those of the authors only. 

\clearpage
\bibliographystyle{ACM-Reference-Format}  
\bibliography{ref}  
\end{document}